# NUMERICAL SIMULATION OF 2D ELECTRODYNAMIC PROBLEMS WITH UNSTRUCTURED TRIANGULAR MESHES

D.A. Fadeev[1]

[1] Institute of Applied Physics RAS

email: fadey.d.a@gmail.com

*Abstract*

We present a generalization of standard leap-frog plus Yee mesh approach for Cauchy problem in electrodynamics simulations on unstructured triangulated mesh. The presented approach still inherits from finite-difference time-domain (FDTD) and do not use techniques developed in finite-volume time-domain approach (FVTD). In the paper the whole flow from mesh creation to actual simulation is presented. The proposed computation flow is parallel ready and can be implemented for distributed systems (computation servers, graphical processing units, etc.). We studied the influence of non-regular triangulation on stability and dispersion properties of numerical solution.

*Introduction*

Typically dynamics of electromagnetic fields in vacuum or media is calculated via advancing of electromagnetic field values set in the nodes of regular Manhattan Yee[1] mesh according to Maxwell equations plus equation for media response. In some cases the above approach is reduced to some scalar equations[2], lower order with respect to time derivative[3, 4, 5] etc. In this paper we will discuss those cases when vector nature of electrodynamic radiation matters and pure Maxwell equations should be solved[6, 7]. In the case of free space or interaction of electromagnetic radiation with soft-bounded media e.g. air plasma[8] regular Manhattan meshes seems to work fine if resolution is enough for media gradients and wavelengths under consideration. In the cases when electromagnetic radiation interacts with condensed media the geometry starts to play a significant role. Even if boundaries are straight lines (e.g. triangle particle) but those lines does not match mesh constraints some local amplification of fields instantly occurs along whole particle surface. These amplifications can be treated in average and for linear problems does not play a significant role. In cases of non-linear problems such amplifications can lead to over estimation of nonlinear terms or/and result in some instabilities. In such cases the general type of meshes have to be used to match the geometry of the objects in simulation area. The most natural mesh is triangulated mesh.

In the next section we will discuss the implementation of algorithm that can advance field values to simulate dynamics of electromagnetic radiation. Then we will study properties of the proposed method. In the last section we will describe how the mesh can be generated and optimized for parallel calculations.

Before getting to the details let us to present basic idea of switching from regular quad meshes to triangulated meshes. Standard Yee mesh shown in fig. 1a assumes setting electrodynamic fields in some specially shifted positions. This allows using a simplest form of discrete approximation of partial derivatives. In Maxwell equations and linear hydrodynamic equations all partial derivatives are of the first order so the best approximation is achieved exactly in middle point with respect to the nodes of operator operands. Very similar thing can be naively achieved if we bind all **E** and **j** vectors to the edges of the triangle mesh, magnetic field **B** vector will be then bound to some center point of triangles. For plasma like media we can attach charge density n to the nodes of the mesh (see Fig.1).

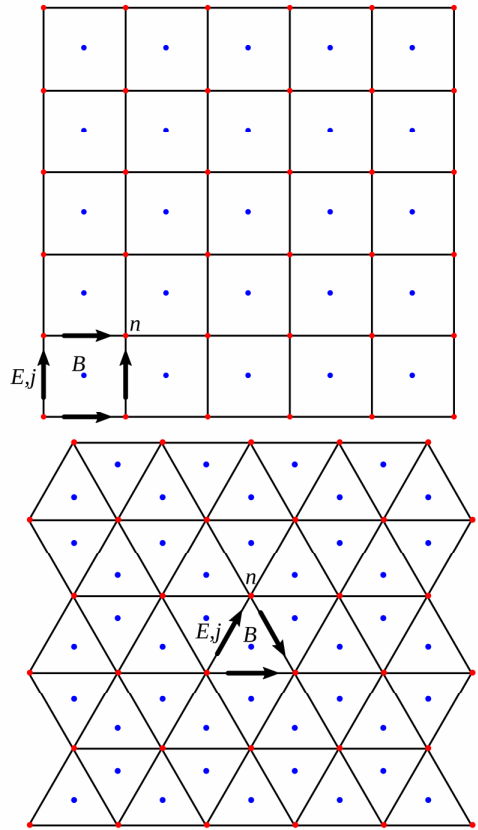

*Fig.1. Generalization of quad mesh (top inset) to triangle mesh (bottom inset) for electrodynamics simulations*

Note that the above approach differs from another known approach utilizing benefits of unstructured meshes FVTD[9, 10]. In that approach both field values **E** and **B** per current cell are bound to same cell point (typically some center point) and both values are known at the same moments of time. In our methodology the field values are known in different time points shifted by $dt/2$ allowing us to adopt leap-frog approach.

## 1. Numerical simulation

We will consider Maxwell equations in vacuum, generalization to plasma like media will be given in the end of this section. The discreet formulation for unstructured mesh remains its general form:

$$\frac{\partial \mathbf{E}}{\partial t} = \hat{\mathbf{L}}_B \mathbf{B}$$
$$\frac{\partial \mathbf{B}}{\partial t} = \hat{\mathbf{L}}_E \mathbf{E} \quad , \qquad (1)$$

Here linear operators $\hat{\mathbf{L}}_E$ and $\hat{\mathbf{L}}_B$ are discreet formulations of curl operator. It is worth to introduce the following short notation for above system as well:

$$\frac{\partial \mathbf{V}}{\partial t} = \hat{\mathbf{M}} \mathbf{V}, \quad \mathbf{V} = \begin{pmatrix} \mathbf{B} \\ \mathbf{E} \end{pmatrix}, \quad \hat{\mathbf{M}} = \begin{pmatrix} 0 & \hat{\mathbf{L}}_B \\ \hat{\mathbf{L}}_E & 0 \end{pmatrix}, \qquad (2)$$

Let us accurately write down the exact form of $\hat{\mathbf{L}}$ operators on triangular mesh. In the Fig. 2 we present all necessary symbols to compose those operators.

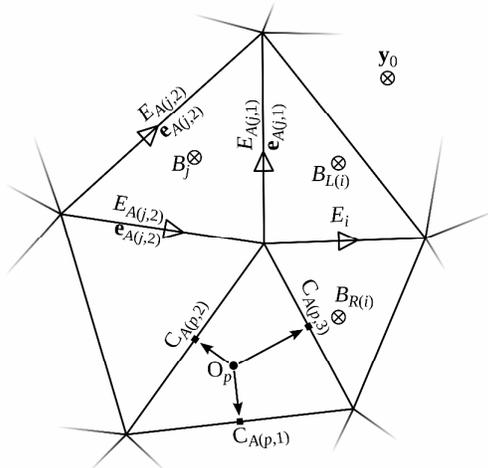

*Fig.2. $E_i$ is the projection of electric field to i-th edge in the center of the edge, $B_i$ is a single magnetic field component (along $\mathbf{y}_0$ orth) in the center of circumcircle of i-th triangle, $\mathbf{e}_i$ is an edge vector with arbitrary direction(since edges are shared we cannot apply any winding rule to choose some specific direction for edge vectors)*

Using notation from Fig 2 operators $\hat{\mathbf{L}}_E$ and $\hat{\mathbf{L}}_B$ will obtain the following form:

$$\frac{\partial E_i}{\partial t} = \frac{B_{R(i)} - B_{L(i)}}{O_{R(i)} O_{L(i)}}, \qquad (3)$$

$$\frac{\partial B_j}{\partial t} = \frac{-1}{S_j} \sum_{q=1}^{3} \mathrm{sgn}\left(\mathbf{y}_0 \left(\overrightarrow{O_j C}_{A(j,q)} \times \mathbf{e}_{A(j,q)}\right)\right) \left|\mathbf{e}_{A(j,q)}\right| E_{A(j,q)}$$
$$, \qquad (4)$$

Here $R(i)$ and $L(i)$ are the indexes of right and left triangles correspondingly with respect to i-th edge. $A(i, j)$ is the index of j-th edge adjacent to i-th triangle. $O_i$ is a circumcircle center of i-th triangle and $C_i$ is the center point of i-th edge.

It can be noted that at least $\hat{\mathbf{L}}_B$ approximation does not achieve its best at the point where $\mathbf{E}$ vector is set since $\mathbf{E}$ is not necessary falls exactly between $O_R$ and $O_L$. The best approximation is achieved for mesh of equilateral triangles. We will discuss meshing and quality criteria two sections later.

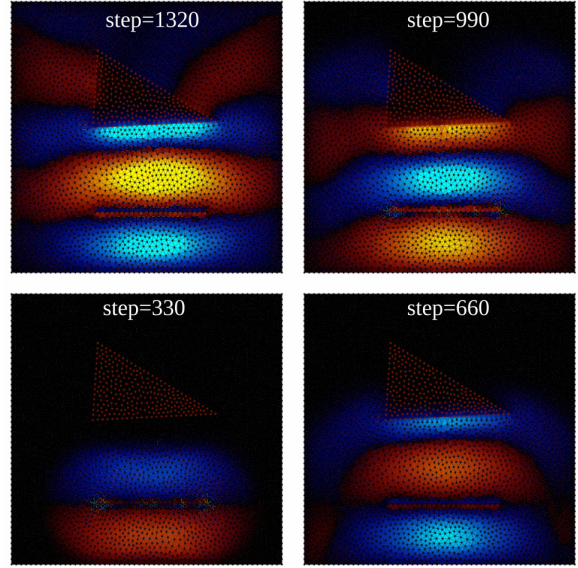

*Fig.3. Snapshots of electromagnetic wave dynamics in the presence of plasma-like object*

It is worth to note that like in structured quad mesh from (3) it follows that div$\mathbf{E}$ remains constant over time. We will not prove this fact in details in order to not overload the scheme in Fig 2 with additional notations. We will just notice that div can be calculated as a flux of $\mathbf{E}$ through the boundary of the Voronoy[11] diagram cell.

To generalize such an approach to plasma like media one can add current vector j to each center of mesh edge. In that case equations for currents become local and without any matrix formalism can be written in the following form:

$$\frac{\partial \mathbf{j}}{\partial t} = n\mathbf{E} - \nu \mathbf{j}, \qquad (5)$$

Since electromagnetic field $\mathbf{E}$ and current density $\mathbf{j}$ are set in the same point and are parallel so no extra interpolation is needed. According to Maxwell equations current should be added to eq.(3) as:

$$\frac{\partial E_i}{\partial t} = \frac{B_{R(i)} - B_{L(i)}}{O_{R(i)} O_{L(i)}} + j_i, \qquad (6)$$

Here we used dimensionless equations, thus no speed of light, electric charge or $4\pi$ factor appears in above equations. This natural generalization of vacuum problem (3, 4) gives the system (4, 5, 6) that can be solved with leap-frog method[12]. In the Fig 3 we present snapshots of solution achieved for a resonator having plasma like triangle. As seen from presented snapshots the quasi planar wave hits the triangle and reflects on it. Solution is stable with criteria for *dt* discussed below. The source code can be found in GitHub page (see the link in the end of the paper).

In the next chapters we will carefully study how switching from structured quad mesh to unstructured triangle mesh impacts dispersion and stability.

### 2. Solution properties

In this section we will study how switching from classic Yee mesh[1] to triangle based unstructured mesh affects stability of solution and it's dispersion properties. First we will discuss the dispersion properties for the discreet formulation of Maxwell PDEs on unstructured mesh. We will start from energy conservation law which is preserved for discreet approximation of operator $\hat{\mathbf{M}}$. Using this relation we will derive some properties of exact solution for discreet problem (i.e. with continuous time). Then we will discuss stability of leap-frog type of solution for Cauchy problem with finite time step *dt*.

*Energy momentum*

To calculate full electromagnetic energy $W = \int(|E|^2 + |B|^2)dS$ one should choose the specific form of scalar product for vector **V**. We will introduce the following scalar product:

$$\mathbf{V}^a \mathbf{V}^b = \sum_{i=1}^{n_e} E_i^a E_i^b \left(O_{R(i)} O_{L(i)}\right)|\mathbf{e}_i| + \sum_{j=1}^{n_t} B_j^a B_j^b S_j, \quad (7)$$

Where $n_e$ and $n_t$ are the number of edges and triangles correspondingly. From the chosen metrics it follows that electromagnetic energy momentum *W* can be calculated as:

$$W = \sum_{i=1}^{n_e} E_i^2 \left(O_{R(i)} C_i + O_{L(i)} C_i\right)|\mathbf{e}_i| + \sum_{j=1}^{n_t} B_j^2 S_j, \quad (8)$$

Using equations (1, 2) for the first derivatives of *E* and *B* the first derivative of the above quantity can be rewritten as:

$$\frac{1}{2}\frac{\partial W}{\partial t} = -\sum_{i=1}^{n_e}(B_{L(i)} - B_{R(i)})E_i|\mathbf{e}_i| + \\ + \sum_{j=1}^{n_t}\sum_{q=1}^{3}\mathrm{sgn}\left(\mathbf{y}_0 \overrightarrow{O_j C_{A(j,q)}} \times \mathbf{e}_{A(j,q)}\right)|\mathbf{e}_{A(j,q)}|E_{A(j,q)} B_j, \quad (9)$$

Last member of right hand side of the above equality can be rewritten in 'per edge' form instead of the above 'per triangle' notation. In such a new notation it is easier to find discreet analog of energy conservation law. We will also assume that electric field set for outer edge centers is zero which correspond to ideal resonator case.

$$\frac{1}{2}\frac{\partial W}{\partial t} = -\sum_{i=1}^{n_e}(B_{L(i)} - B_{R(i)})E_i|\mathbf{e}_i| - \\ -\sum_{i=1}^{n_e}\left(\mathrm{sgn}\left(\mathbf{y}_0 \overrightarrow{O_{R(i)} C_i} \times \mathbf{e}_i\right)\right)B_{R(i)} + \\ +\mathrm{sgn}\left(\mathbf{y}_0 \overrightarrow{O_{L(i)} C_i} \times \mathbf{e}_i\right)B_{L(i)}\right)E_i|\mathbf{e}_i| = , \quad (10) \\ = -\sum_{i=1}^{n_e}(B_{L(i)} - B_{R(i)})E_i|\mathbf{e}_i| - \\ -\sum_{i=1}^{n_e}(B_{R(i)} - B_{L(i)})E_i|\mathbf{e}_i| = 0$$

More accurately the above equation can be rewritten in a divergent form for Poynting vector, but for the sake of simplicity we restricted ourselves with ideal resonator case with boundary conditions $E_\Sigma = 0$, where $\Sigma$ is any boundary edge index.

From discreet energy conservation law(10) it follows that
- all eigenvectors of M operator are real;
- every two eigenvectors with different eigenvalues are orthogonal in the scope of metric (7).

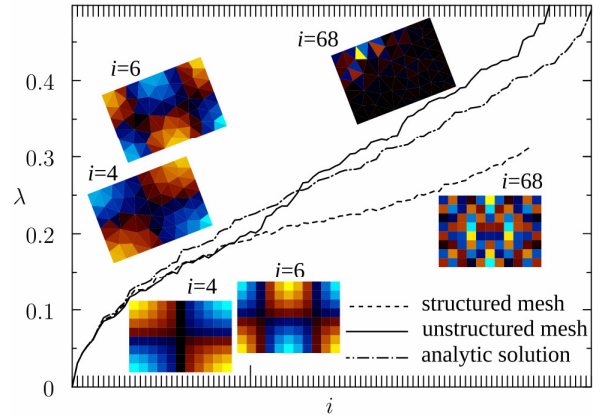

*Fig.4. Eigenvalues against mode (eigenvector) index for rectangular resonator with dimensions 70 × 101. Comparison of structured quad mesh against unstructured triangle mesh and exact analytic solution (see markers in legend). Modes are shown as colormap insets plotted for magnetic field value. Tilted colormaps are for unstructured mesh, straight ones are for structured mesh; mode number is specified near each inset*

With the above conclusions we can use "complex time" method to find eigenvectors and eigenvalues. The idea is based on the fact that if t is substituted by it new eigenvalues will become real, which does actually follow from energy conservation law (10). First eigenvector with maximal eigenvalue can be then found by updating new system with either Euler or leap-frog method. Being normalized at each step at some point any random initial vector will converge to first eigenvector since all other solutions will not survive due to lower eigenvalues. Next eigenvectors can be found using the fact that all eigenvectors with different eigenvalues are orthogonal in metric (7). So we can withdraw found eigenvector from any initial vector and start from it. By removing first eigenvector at each step we then will not allow this dominant eigenvector to rise from round-off errors, and finally will come to next eigenvector. Complete solution can be found in GitHub (see the link in the end of the paper).

In the Fig 4 we plotted eigenvalues for all electromagnetic modes of rectangular resonator. The number of that eigenvalues equal to the number or cells in mesh or number of B vector components. Note that all eigenvalues has their copies with negative sine. We also has a number of zero eigenvalues describing eigen-subspace of static solutions. From fig. 4 it is well seen that for lowest eigenvectors all three unstructured, structured and analytic solutions are quite close, then analytic solution starts to show a difference while structured and unstructured ones remains quite close. This is a natural behavior for discreet problems related to so-called numerical dispersion. Then unstructured solution suddenly drops of the structured one. This happens due to completely different type of eigenvectors for structured and unstructured triangle meshes (compare colormap insets for mode number 68).

The main conclusion here is that for eigenvectors that barely correspond to normal modes of ideal resonator structured and unstructured discretization give very close results. We will get back to fig 4 right away after introducing a stability criteria for leap-frog scheme.

### *Stability*

To study stability of proposed numerical scheme we will study stability of solution for modes discussed in previous section. Briefly speaking we will generalize well known leap-frog method stability criteria for oscillatory motion[12]. First we will write the system of equations for *i*-th mode, having eigenvalue $\lambda_i$:

$$\frac{\partial \mathbf{V}_i}{\partial t} = \hat{\mathbf{M}} \mathbf{V}_i = \lambda_i \mathbf{V}_i, \quad (11)$$

$$\frac{\partial \mathbf{B}_i}{\partial t} = \hat{\mathbf{L}}_B \mathbf{B}_i = \lambda_i \mathbf{E}_i = i\omega_i \mathbf{E}_i, \quad (12)$$

$$\frac{\partial \mathbf{E}_i}{\partial t} = \hat{\mathbf{L}}_E \mathbf{E}_i = \lambda_i \mathbf{B}_i = i\omega_i \mathbf{B}_i. \quad (13)$$

Now by applying leap-frog method we will achieve the following (further *i* is omitted):

$$\mathbf{E}(t+dt) = \mathbf{E}(t) + \hat{\mathbf{L}}_B \mathbf{B}\left(t+\frac{dt}{2}\right)dt =$$
$$= \mathbf{E}(t) + i\omega \mathbf{E}\left(t+\frac{dt}{2}\right)dt \quad (14)$$

$$\mathbf{B}\left(t+\frac{3dt}{2}\right) = \mathbf{B}\left(t+\frac{dt}{2}\right) + \hat{\mathbf{L}}_E \mathbf{E}(t+dt)dt =$$
$$= \mathbf{B}\left(t+\frac{3dt}{2}\right) + i\omega \mathbf{B}(t+dt)dt \quad (15)$$

By substituting the equations to each other we are making sure that for both E and B sub-vectors we have the same equation. Now let us consider this equation for any vector **X**:

$$\mathbf{X}(t+dt) = \mathbf{X}(t) + i\omega \mathbf{X}\left(t+\frac{dt}{2}\right)dt. \quad (16)$$

Let us study solution properties by substituting exponential solution $\mathbf{A}_0 \exp(i\omega(t+dt))$:

$$\mathbf{A}_0 \exp(i\omega(t+dt)) = \mathbf{A}_0 \exp(i\omega t) +$$
$$+ i\omega \mathbf{A}_0 \exp\left(i\omega\left(t+\frac{dt}{2}\right)\right)dt \quad , \quad (17)$$

$$\exp(i\omega dt) = 1 + i\omega \exp\left(\frac{i\omega dt}{2}\right)dt, \quad (18)$$

$$\exp\left(\frac{i\omega dt}{2}\right) = \frac{i\omega dt}{2} \pm \sqrt{1 - \frac{\omega^2 dt^2}{4}}. \quad (19)$$

Note that in above expression if the square root is real argument of exponent can be treated as pure complex number. So we come to stability criteria in the following form:

$$\omega \in \Re \quad \text{if} \quad \omega dt < 2. \quad (20)$$

Stability is achieved if for every mode the above criterion is fulfilled. So for whole scheme stability is achieved if dt < 2/$\omega_{max}$, where $\omega_{max}$ is the cyclic frequency of highest mode. Obviously the criterion (20) is also true for structured mesh. Getting back to Fig 3 it is well seen that unstructured mesh is a little more unstable than structured one and requites smaller time steps. But at the same time it has less electric field components: 3/2 per magnetic field component compared to 2 in case of quad structured mesh.

### *3. Mesh generation*

In order to create a mesh for simulation one need to seed geometry i.e. add points to contours of boundary, objects and possibly some paths. Then using same seeding parameter one should add some volume(area in 2d) nodes to mesh. For this node cloud classical Delaunay method [13] can be used to make triangles covering the convex hull of node cloud.

The first problem of above scheme is that generated triangles can cross the geometry. As non-robust solution we use 'flip' refinement method that flips pair of triangles contacting the edge that crosses the geometry (see Fig 5 a). This method has a drawback when multiple adjacent triangles cross the geometry edge.

The next problem is seeding of volume. This is expected to create a uniform fill but having some predefined geometry we can produce very small triangles. According to our results from previous chapter this is not acceptable for leap-frog scheme. Each tiny triangle will produce a mode with very high oscillatory frequency forcing us to lower *dt* and loose performance. To overcome the above problem we can run an iterative process of mesh refinement which will shift nodes in a way to make nodes that are far from each other to become closer and vice versa. This can be done by solving a physical problem of nodes connected by springs along triangle edges and along heights (see fig 5 b). The initial spring length can be cal-culated in assumption that we magically generated mesh of perfect triangles covering whole volume. This will make uniform meshing that is needed in most cases for Maxwell equations simulations. If dense meshing is needed in some area springs length should become a field variable over space. Since we want a static solution we need to introduce quite big dumping to nodes motion. By neglecting second order derivatives over time we also can simplify the problem of relaxation. This will lead as to jello model which can be easily solved either with Euler method. After some iterations of the above jello model we need to remesh the node cloud because some triangles may become significantly deformed and cancel further relaxation in their neighboring area. Plus we need to anchor seeding nodes of input geometry.

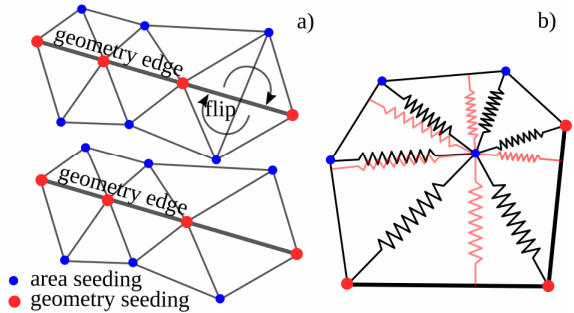

*Fig.5. Meshing algorithms. a) – flip operation, b) – jello model for area seeding*

Later on some triangles of convex hull can be removed to cover just needed area. We typically add bounding box around input geometry and then remove all elements that are out of input geometry. The above algorithm was implemented and can be found in GitHub (see the link in the end of article).

### *Divide and conqueror*

It is well known that classic leap-frog plus Yee mesh approach is highly local so it can be easily divided among calculating units and processed in parallel. The same is also true for an approach with unstructured meshes. The simplest idea of partitioning is to use KD tree idea. So having the depth of partitioning at each step one divide each part along x or y direction by two parts with same amount of elements. This works well for areas with rectangular boundary and produce partitioning with optimal shared boundary length. At the same time by the algorithm each part has nearly identical number of elements i.e. number of elements deviation is small.

We have to fulfill both deviation and shared boundary criteria. Since we cannot use queue due to data transfer overhead we need to balance the load for each compute unit. It is also evident that the bigger partition is the smaller shared boundary size is, which reduces amount of data shared between compute units at each step.

For tricky geometries 'KD' partitioning may cross geometry boundary contour some bad way and produce uncontrollable small partitions. This can be solved with 'greedy' partitioning method. This method takes any element in the mesh and grows the area by adding neighboring elements at each step. When number of elements reaches desired number the algorithm switches to another partition by choosing another root node that was not 'eaten' before. It was found that such an algorithm performs best if each partition is started from the 'sharpest' node i.e. the node on existing boundary that has smallest number of neighbors.

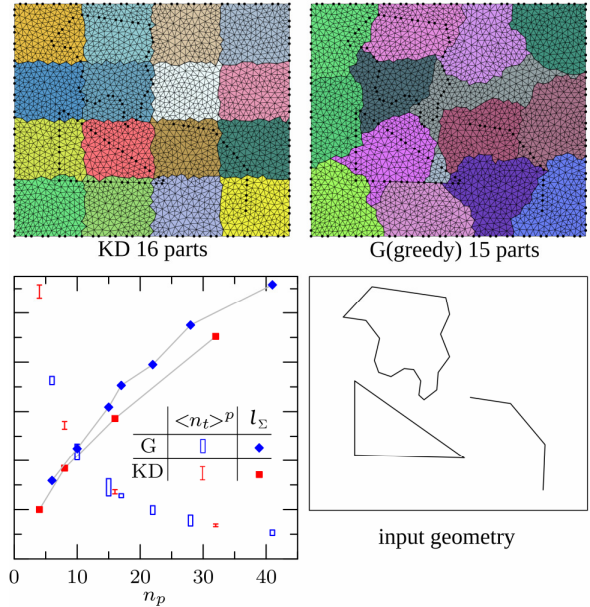

*Fig.6. Comparison of partitioning algorithms. Top tiles show partitioning examples with nearly the same number of parts with KD(left) and 'greedy' (right). In the lower right tile input geometry is shown. The plot shows overall shared boundary length $l_\Sigma$ i.e. number of shared edges and average number of triangles per part $<n_t>^p$ for KD and 'greedy' methods, error bars show root mean square deviation*

After the partitioning is done with either KD or 'greedy' the result can be refined. We used two types of refinement:

- Join very small(smaller than threshold) parts with neighboring parts. We join to smallest part above with size the threshold.
- Join 'hanging' triangles to neighboring part. 'Hanging' are the triangles that contact own part by only one edge and with other two edges contact single neighboring part. Triangles contacting three parts are not refined.

In Fig 6 we present partitioning result with all refinements are done. From the fig. 6 it is seen that as expected 'KD' produces more stable parts in terms of size deviations. The shared boundary size is nearly the same. We can conclude that 'greedy' is not so bad compared to KD, having certain important advantages over KD:

- 'greedy' not sensitive to non-trivial geometries
- KD produces 2n parts, 'greedy' can produce any number of parts.

We hope that partitioning will allow to use all the above methodology with any distributed system from classical clusters to GPU or Intel Phi clusters.

*Conclusion*

Using unstructured meshes seems to be quite interesting for modern physics applications dealing with sophisticated geometries like nano-structured surfaces and particle arrays. Numerical methods for unstructured meshes appear to be a natural generalization of methods used for structured meshes, preserving their important features. Simulations with unstructured meshes seems to be easy adoptable for modern computational systems. In this article we studied 2d case for easiest linear system of Maxwell equations. Methods used in this work can be generalized to 3d tetrahedral meshes, while components of electromagnetic fields, currents and plasma density can be still bind to edges, faces and nodes correspondingly.

In many practical applications on the boundary of mesh PML layers are adopted to simulate open boundary. The proposed method can work with mixed type of mesh e.g. triangular plus quad. So the meshes having rectangular boundaries can be tied with structured rectangular frame shaped mesh with PML layers.

All numerical codes used in this article are open source and located at GitHub:
https://github.com/dafadey/hed.git
A full movie for Fig 3 can be obtained at
https://www.youtube.com/watch?v=SryTESknKDE

*Acknowledgements*

Author acknowledges the support from Federal Research Center Institute of Applied Physics of the Russian Academy of Sciences (project No.0035-2014-0020) and program of the Presidium of the Russian Academy of Sciences "Nonlinear dynamics in mathematical and physical sciences" (project No 0035-2018-0006).